\documentclass[prl,letterpaper,twocolumn,groupedaddress,amsmath,amssymb,showpacs]{revtex4}

\usepackage{graphicx}
\usepackage{bm,mathrsfs}
\begin{document}

\title{Strong-field interactions between a nanomagnet and a photonic cavity}
\author{\"Oney O. Soykal}
\author{M. E. Flatt\'e}
\affiliation{
Optical Science and Technology Center and Department of Physics and Astronomy\\ 
University of Iowa, Iowa City, IA 52242}


\begin{abstract}

We analyze the interaction of a nanomagnet with a single photonic mode of a microcavity in a fully quantum-mechanical treatment and find 
that exceptionally large quantum-coherent magnet-photon coupling can be achieved. Coupling terms in excess of several THz are predicted to be achievable in a spherical cavity of $\sim 1$~mm radius with a nanomagnet of $\sim 100$~nm radius and ferromagnetic resonance frequency of $\sim 200$~GHz.  Eigenstates of the magnet-photon system correspond to entangled states of spin orientation and photon number, in which over $10^5$ values of each quantum number are represented; conversely initial (coherent) states of definite spin and photon number evolve dynamically to produce  large oscillations in the microwave power (and nanomagnet spin orientation), and are characterized by exceptionally long dephasing times.

\end{abstract}


\maketitle


Strong coupling between light and electronic transitions \cite{Dicke,Cummings,JCM,Kimble1998,Raimond2001,Kiraz2001,Walther2002} permits coherent transfer of quantum information between the two systems, as well as a host of exotic phenomena, including slow light \cite{Imamoglu,Harris}, lasing without population inversion \cite{Scully-p,Padmabandu}, and index enhancement via quantum coherence \cite{Mandel,Sultana}. Achieving strong coupling between light and electronic transitions in solids has been more challenging, due to the shorter coherence time of electrical dipole transitions in solids compared to atoms, however, strong coupling in a single quantum dot-semiconductor microcavity system \cite{Forchel} has been demonstrated with a coupling strength  \hbox{$\sim 80$~$\mu$eV}. Often these investigations in solids focus on electric dipole (orbital) transitions over magnetic (spin) transitions, whose typical oscillator strengths are estimated\cite{Jackson2} to be smaller by a factor of the fine structure constant, \hbox{$\sim 1/137$}. Paramagnetic spin systems in solids, however, appear intrinsically more quantum coherent than orbital coherent states\cite{Awschalom2002,Awschalom2007}, and collective spin-photon effects (such as superradiance\cite{Feher1958,Benedict1996}, including in molecular magnets of $\sim 10$ spins\cite{Chudnovsky2002}) have also been explored. Yet to be explored are the coherent strong-field properties of ferromagnetic systems. In ferromagnets the exchange interaction can cause a very large number of spins to lock together into one macrospin with a corresponding increase in oscillator strength. Therefore for nanomagnets with more than $\sim 100$ spins, the electronic-photonic coupling strength may exceed that of a  two-level electronic orbital transition occurring by electric dipole coupling, while still maintaining the long coherence times (ferromagnetic nanomagnet oscillators have been demonstrated\cite{Rippard,Sankey2006} with Q factors in excess of $500$). Such ferromagnetic oscillations can be coherently driven by electrical spin currents\cite{Berger,Slonczewski,Buhrman,Ralph,Sankey2006,Bass}, and thus a single nanomagnet-photonic mode system provides an efficient method of strongly coupling electronic, magnetic and photonic degrees of freedom.
 
\begin{figure}[htp]
  \centering
  \includegraphics[width=6.0 cm]{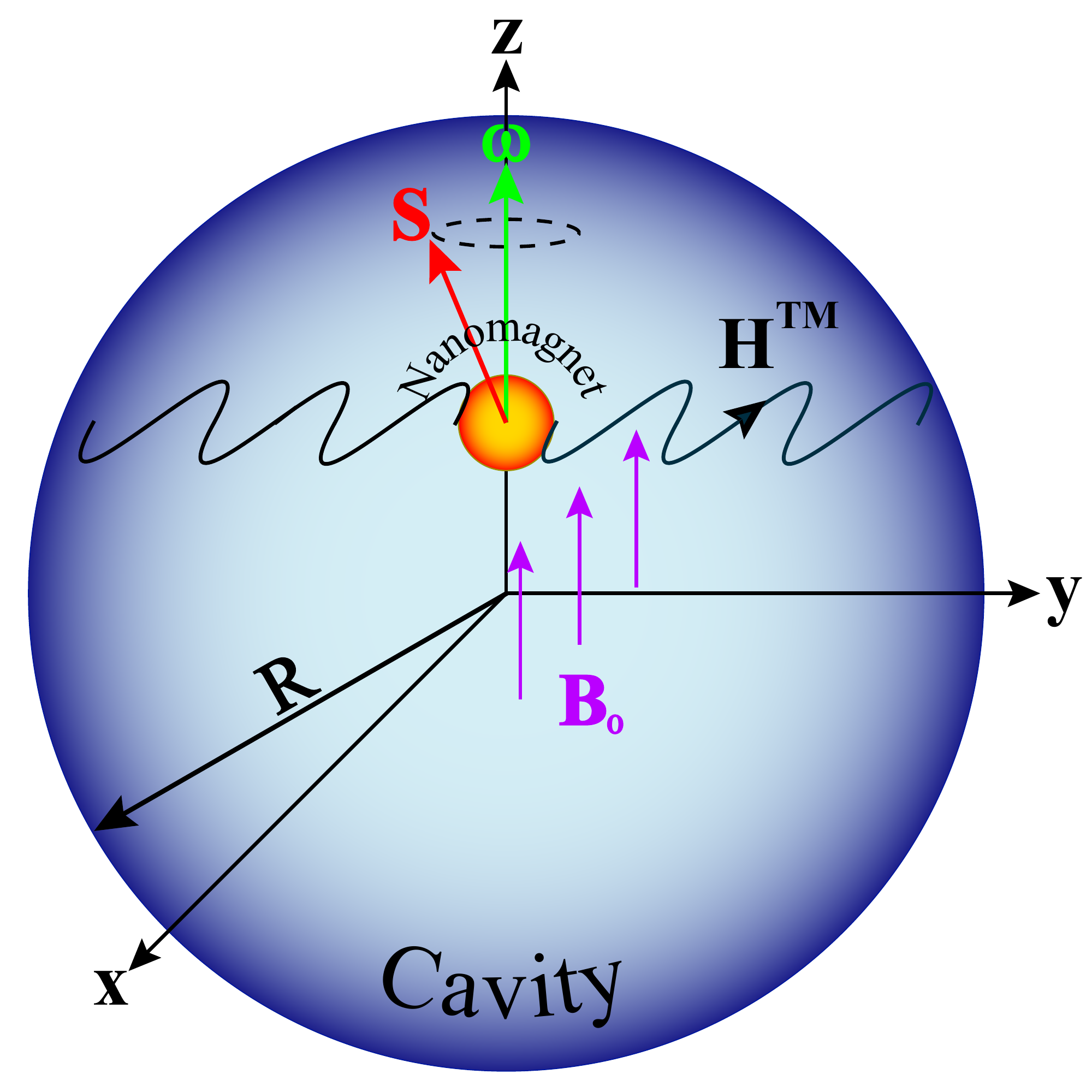}
  \caption{Schematic of nanomagnet-microcavity system with a spherical nanomagnet of radius $r_m$ placed at a distance of $d$ from the center of a microcavity of radius $R$. A uniform magnetic field, $\bm{B}_0$, is applied along the $\mathbf{z}$-axis causing precession of the nanomagnet macrospin, $\bm{S}$, with frequency of $\omega$, in resonance with a photon mode of the cavity.}\label{fig:schematic}
\end{figure}
Here we calculate the strong-field interactions between a small ferromagnet (nanomagnet) and light, and find a dramatic enhancement of spin-photon coupling relative to paramagnetic spin systems, yielding coupling much larger than found by coupling light to orbital transitions.   As shown schematically in Fig.~\ref{fig:schematic}, the oscillator is a spherical nanomagnet with a radius $r_m$, possessing $N$ spins (for $r_m\sim 100$nm, $N\sim 10^9$), placed a distance $d$ from the center of the cavity for more efficient coupling to the cavity mode. High frequency precession of the nanomagnet at a frequency resonant with the cavity is achieved by tuning a uniform magnetic field $\bm{B}_0$ along the $\mathbf{z}$-axis of the cavity. We find for a realistic cavity size ($\sim 1$~mm) and resonance frequency ($\sim 200$~GHz) that photonic coupling terms between neighboring spin states when the cavity is empty of photons are comparable in size to those of quantum dot electric dipole transitions ($\sim 100$~$\mu$eV)\cite{Forchel}. We also find an unexpected regime, however, in which the system is initialized with (1) no photons and (2) the nanomagnet in its high-energy (antiparallel) orientation to the magnetic field, whereby large oscillations in spin number and photon number result, corresponding to spin-photon coupling strengths between neighboring spins $\sim 160$~meV ($\sim 40$~THz). These large oscillations in spin and photon number ($\sim 10^5$ quanta of each) on $\mu$s timescales are characterized by long dephasing times.

The total Hamiltonian of the system incorporates the magnetic $\bm{H}$ and electric fields $\bm{E}$ of the cavity and the magnetization $\bm{M}$ of the nanomagnet\cite{Jackson},
\begin{equation}
\mathscr{H}=\frac{1}{2}\int\left(\mu_0|\bm{H}|^2+\epsilon_0|\bm{E}|^2+\mu_0
\left(\bm{H}\cdot\bm{M}\right)\right)d^3r.\label{1}
\end{equation}
 The first two integrands on the right hand side of Eq.~(\ref{1}) correspond to the free field Hamiltonian, whereas the third integrand describes the interaction Hamiltonian of the nanomagnet-cavity system. Spherical wave expansion of the cavity field \cite{Heitler}, and renormalization of the respective field strength coefficients to satisfy the Weyl-Heisenberg commutation relations, $[a_{lm},a_{l^\prime m^\prime}^\dagger]=\delta_{ll^\prime}\delta_{mm^\prime}$, yields the following form of the interaction Hamiltonian
\begin{eqnarray}
\mathscr{H}_I&=&\left[\sum_{l,m}\Gamma_{l}^{(TE)} a_{lm}^{(TE)}\int_S\bm{M}\cdot\left(\bm{\nabla}\times\bm{u}_{lm}\right)d^3r\right.\nonumber \\ &&+\left.\sum_{l,m}\Gamma_{l}^{(TM)} a_{lm}^{(TM)}\int_S\bm{M}\cdot\bm{u}_{lm}\,d^3r \right]+ c.c.\label{2}
\end{eqnarray}
where the appropriate basis functions for spherical waves are given by $\bm{u}_{lm}$. Moreover, the coupling constants for transverse electric, and transverse magnetic modes of the field for angular momentum $l$ are $\Gamma_l^{(TE)}$ and $\Gamma_l^{(TM)}$, respectively.

A nanomagnet acting as a macrospin, as seen experimentally in nanomagnet oscillators of roughly this size\cite{Sankey2006}, has a magnetization
\begin{equation}
\bm{M}=\bm{\mu}/V=-\frac{g_s\mu_B}{\hbar V}\bm{S}\Theta(r_m-|\bm{r}-\bm{d}|),\label{3} 
\end{equation}
in terms of the collective spin operator $\bm{S}$ and the Heavyside step function $\Theta(x)$. A spherical nanomagnet of radius $r_m \cong 108$ nm, consisting of iron (magnetic moment $2.21\mu_B$ per atom), possesses $N\sim 10^9$ spins. 

The spherical wave expansion of the magnetic field has several unique features such that all components of the field are identically zero if $l=m=0$, dictating that there are no radiating monopoles. For a magnetic field applied in the $\hat z$ direction, the microwave emission of the nanomagnet is due to the oscillating components of the magnetization $\bm{M}_{x,y}$ perpendicular to the radial direction (Fig.~\ref{fig:schematic}), and thus the cavity TE mode (which has a magnetic field pointing in the radial direction) does not couple to the nanomagnet. Therefore the nanomagnet will only be coupled to TM cavity photons. The appropriate basis functions for the lowest-frequency (and assumed dominant) dipole TM mode ($l=1$) are given by $\bm{u}_{1m}=g_1(kr)\mathbf{Y}_{1,1,m}(\theta,\phi)$ in terms of the spherical Bessel functions and the vector spherical harmonics.  A vector spherical harmonic expansion can be written in the helicity basis, using the helicity basis vectors $\mathbf{\hat{e}_m}$ (which form a spherical tensor of rank $1$ ($m=0,\pm 1$)) with field strength coefficients $a_{1,\pm 1}$ ($a_{1,\pm 1}^*$).  The spin operator of the nanomagnet, written also in the helicity basis,
\begin{equation}
\bm{S}=\frac{1}{\sqrt{2}}(S_+\mathbf{\hat{e}_-}-S_-\mathbf{\hat{e}_+})+S_z\mathbf{\hat{e}}_0,\label{4}
\end{equation}
in terms of the nanomagnet spin raising and lowering operators. Introduction of this collective spin operator to Eq.~(\ref{2}), as well as replacing the field strength coefficients of the TM mode with the corresponding annihilation (creation) operators, yields a fully quantum Hamiltonian
\begin{equation}
\mathscr{H}_\gamma=\hbar\omega_\gamma\left(a_\gamma^\dagger a_\gamma+\frac{1}{2}\right)-g\mu_B\Gamma_\gamma
\left(a_\gamma S_+ + a_\gamma^\dagger S_-\right)+g\frac{\mu_B}{\hbar}B_0 S_z,\label{5}
\end{equation}
in which the spin interacts only with a single photon mode $\gamma$. Modes of higher $\ell$ would be out of resonance because of the cavity quantization, and energy non-conserving terms with negative helicity have been dropped (relying on the rotating wave approximation \cite{Scully}). The nanomagnet-photon coupling constant, $\Gamma_\gamma$, is found to be
\begin{equation}
\Gamma_\gamma=\frac{j_1(kd)}{8\hbar|j_1(y_{1\gamma})|}\left[1-\frac{l(l+1)}{y_{1\gamma}^2}\right]^{-1/2}
\sqrt{\frac{3\hbar\omega_\gamma\mu_0}{\pi R^3}},\label{6}
\end{equation}
where $y_{1\gamma}$ is the $\gamma$-th zero of $|rj_1(kr)|^\prime$ satisfying the appropriate conditions for the field for the TM mode at the cavity boundary. The mode frequency $\omega_{\gamma}$ is related to the radius of the cavity $R$ with $k_{1\gamma}=\omega_{1\gamma}/c=y_{1\gamma}/R$.
Furthermore, from Eq.~(\ref{5}) the cavity is in resonance with the energy level splitting of the spins whenever the relation $\hbar\omega = g\mu_B B_0$ is satisfied. Therefore, any spin flip up (down) process of the nanomagnet results in an absorption (emission) of a cavity photon in the case of exact resonance. An applied uniform magnetic field of $B_0 = 7$ T, corresponding to a precession of the macrospin with a frequency of $\sim 200$ GHz, will cause the nanomagnet spins to be in exact resonance with a cavity volume of $1.25$ mm$^3$. We assume the lowest TM mode of the cavity is in resonance with the spin-flip transitions of the nanomagnet, so as higher-energy modes will not be in resonance the subscript $\gamma$ will be dropped from Eq.~(\ref{5}).

The eigenstates of the nanomagnet, treated as a macrospin, are simultaneous eigenstates of the collective spin operators $\bm{S}^2$, and $\bm{S}_z$ given by $|l_s,m_s\rangle$, where $|m_s|\leq l_s\leq N/2$. Part of the macrospin approximation is the assumption that $l_s$ is fixed, and we assume the (most likely) maximal spin state, $l_s = N/2$. The total excitation number 2$\xi$, corresponding to the maximum number of photons $n$ in the cavity (when the nanomagnet is parallel to the static magnetic field), is conserved by the Hamiltonian of Eq.~(\ref{5}). 
For an initial configuration of the macrospin pointing antiparallel to the static field $\bm{B_0}$ and no photons in the cavity, $\xi=N/2$, the basis states of the spin-photon mode system $|n,m_s\rangle$ can be written as  $|n,\xi-n\rangle$ or $|\xi-m_s,m_s\rangle$, so that the basis states are indexed either solely by photon number of the cavity ($n$), or by eigenvalue of $\bm{S}_z$ ($m_s$). The structure of these basis states is similar to those of the Dicke model\cite{Dicke} for $N$ independent atomic spins, wherein $l_s$ is the  \textit{cooperation number} of the paramagnetic collection of spins. This is as it should be, for the Hilbert space of $N$ independent spins should include the states of a macrospin corresponding to $l_s = N/2$. The assumption $\xi=N/2$ corresponds to the initially fully excited atomic system in the Dicke model, with no photons in the cavity.  
In our system, however, elements of the Hilbert space with $l_s\ne N/2$ are split off in energy due to the exchange interaction.

To proceed, we adopt the notation $|n,\xi-n\rangle$ and drop the redundant reference to the $m_s$, so the total Hamiltonian takes the form of
\begin{equation}
\mathscr{H}=\sum_{n=0}^{2\xi}E_0|n\rangle\langle n|-\tau(n)\left[ |n+1\rangle\langle n| + |n\rangle\langle n+1|\right],\label{7}
\end{equation}
in the Fock space, where the constant energy coefficient $E_0$ term and the coupling strength $\tau(x)$ are defined as,
\begin{eqnarray}
E_0&=&\hbar\omega\left( \xi+1/2\right),\nonumber\\
\tau(n)&=&\hbar\Gamma g\mu_B(n+1)\sqrt{2\xi-n}\,\, .\label{8}
\end{eqnarray}
For $2\xi=N$, the magnet-microwave mode coupling, $\tau(n)$, changes over a range of $0.10$ Mhz - $4.1$ THz through all possible photon (spin) numbers.  
 $\tau(n)$ acts like a driving force for a fictitious particle moving between sites  labelled by photon number $n$, so $|0\rangle\rightarrow ...\, ...\rightarrow |n-1\rangle \rightarrow |n\rangle \rightarrow |n+1\rangle\rightarrow ...\, ...\rightarrow |2\xi\rangle$. The solutions $n_o$ of $\tau^\prime(n)|_{n_0}=0$ are equilibrium points in cavity photon number, and for this system there is one, $n_0=(4\xi-1)/3$. The coupling can also be expressed in terms of the collective spin number $m_s$ as $\tau(m_s)=\hbar\Gamma g\mu_B(\xi-m_s+1)\sqrt{\xi+m_s}$, with an equilibrium point of $m_0=(1-\xi)/3$. For a system consisting of a very large number of spins ($\xi\gg 1$), the eigenfunctions of the Hamiltonian in Eq.~(\ref{7}) are expected to be centered about $n_0=4\xi/3$ as well as $m_0=-\xi/3$. 

For an initial state $|n,m_s\rangle$, if we are only interested in transitions which conserve energy and in which a photon is emitted, the rate of photon emission $R_n$ is proportional to $\sum_{\forall \Psi}|\langle\Psi|a^\dagger S_- |n,m_s\rangle |^2$, where $|\Psi\rangle$ represents the possible final states of the system. Therefore, $R_n=A(n+1)^2(2\xi -n)$, or equivalently $R_n=A(\xi -m_s+1)^2 (\xi+m_s)$. The factor $A$ can be identified as the Einstein A-coefficient by applying $R_n$ to a single spin pointing upward ($\xi=m_s=1/2$) when the cavity has no photons ($n=0$). Since $R_n$ reaches its maximum value of $4A(N/3)^3$ for the equilibrium point $m_0$ (or $n_0$) in the large spin limit, the equilibrium points $n_0$ and $m_0$ are the photon number and spin number, respectively where the nanomagnet-cavity system exhibits \textit{superradiance}\cite{Dicke}.
\begin{figure}[htp]
\centering
\includegraphics[width=9 cm]{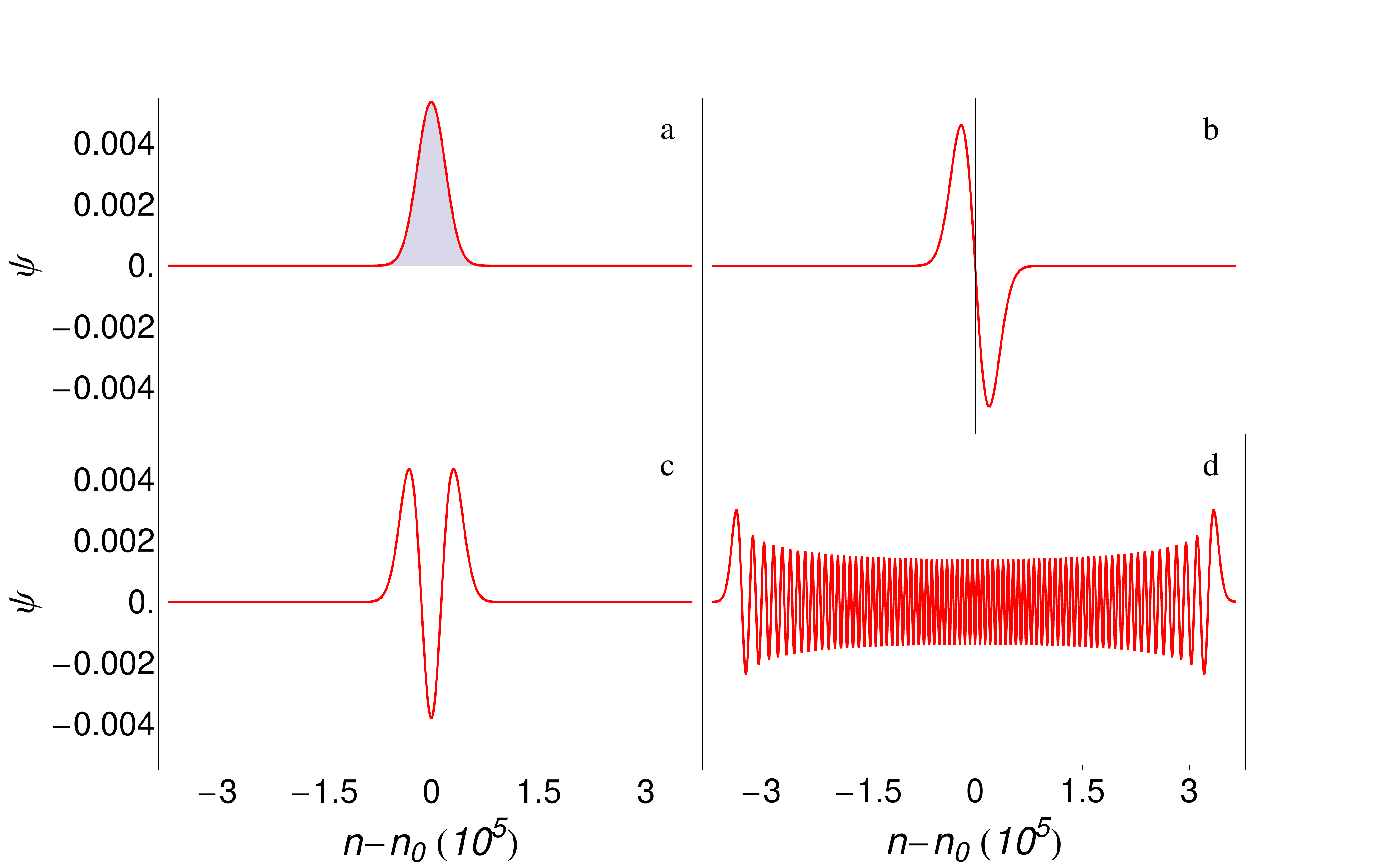}
\caption{Selected wavefunctions of the nanomagnet-cavity system as a function of photon number, $n$, centered about the equilibrium point $n_0=4\xi/3=6.66667\times 10^8$ for $N=10^9$ spins: (a) ground state with a width of roughly $10^5$ photons(spin flips), (b) $1^{st}$, (c) $2^{nd}$, and (d) $150^{th}$ excited states.}\label{fig:wavefunctions}
\end{figure}

The eigenfunctions of the nanomagnet-cavity Hamiltonian given in Eq.~(\ref{7}) can be expanded as $\Psi_j=\sum_{n^\prime}^{2s_z}\psi_j^{n^\prime} |n^\prime\rangle$ in the field basis. Since our nanomagnet consists of very large number of spins ($N=10^9$), the eigenfunctions of the Hamiltonian can be found in the continuum limit, corresponding to replacing $\psi_j^n\rightarrow \psi_j(n\varepsilon)$. Eq.~(\ref{7}) then becomes
\begin{equation}
E_j \psi_j(n\varepsilon)+\tau(n\varepsilon)\psi_j(n\varepsilon+\varepsilon)+\tau(n\varepsilon-\varepsilon)
\psi_j(n\varepsilon-\varepsilon)=0,\label{9}
\end{equation}
which can be  transformed to the following ordinary differential equation
\begin{eqnarray}
&&\tau(x)\frac{d^2\psi_j(x)}{dx^2} + \frac{d\tau(x)}{d x}\frac{d\psi_j(x)}{d x} \label{10}\\
&&+\left( 2\tau(x)-\frac{d\tau(x)}{d x}+\frac{1}{2}\frac{d^2\tau(x)}{d x^2}+E_j\right)\psi_j(x)=0,\nonumber
\end{eqnarray}
with the boundary conditions of $\psi_j(0)=\psi_j(2s_z)=0$  by Taylor expanding the amplitudes $\psi_j(x)$ in Eq.~(\ref{9}) up to the order of $o(\varepsilon^3)$ and defining $n\varepsilon= x$. The lowest energy eigenvalues $E_j$  and eigenfunctions $\psi_j(x)$ of this differential equation, shown in Fig.~\ref{fig:wavefunctions}, can be obtained in the WKB approximation from
\begin{equation}
S(E_j)=\frac{1}{2\pi}\oint\sqrt{\frac{E_j-V_{e}(x)}{\tau(x)}}dx=j+\frac{1}{2},\label{11}
\end{equation}
where the effective potential is given by $V_e(x)=\tau^{\prime}(x)+{\tau^\prime}^2(x)/4\tau(x)-\tau^\prime(x)/2\tau(x)-2\tau(x)$.
\begin{figure}[htp]
\centering
\includegraphics[width=8 cm]{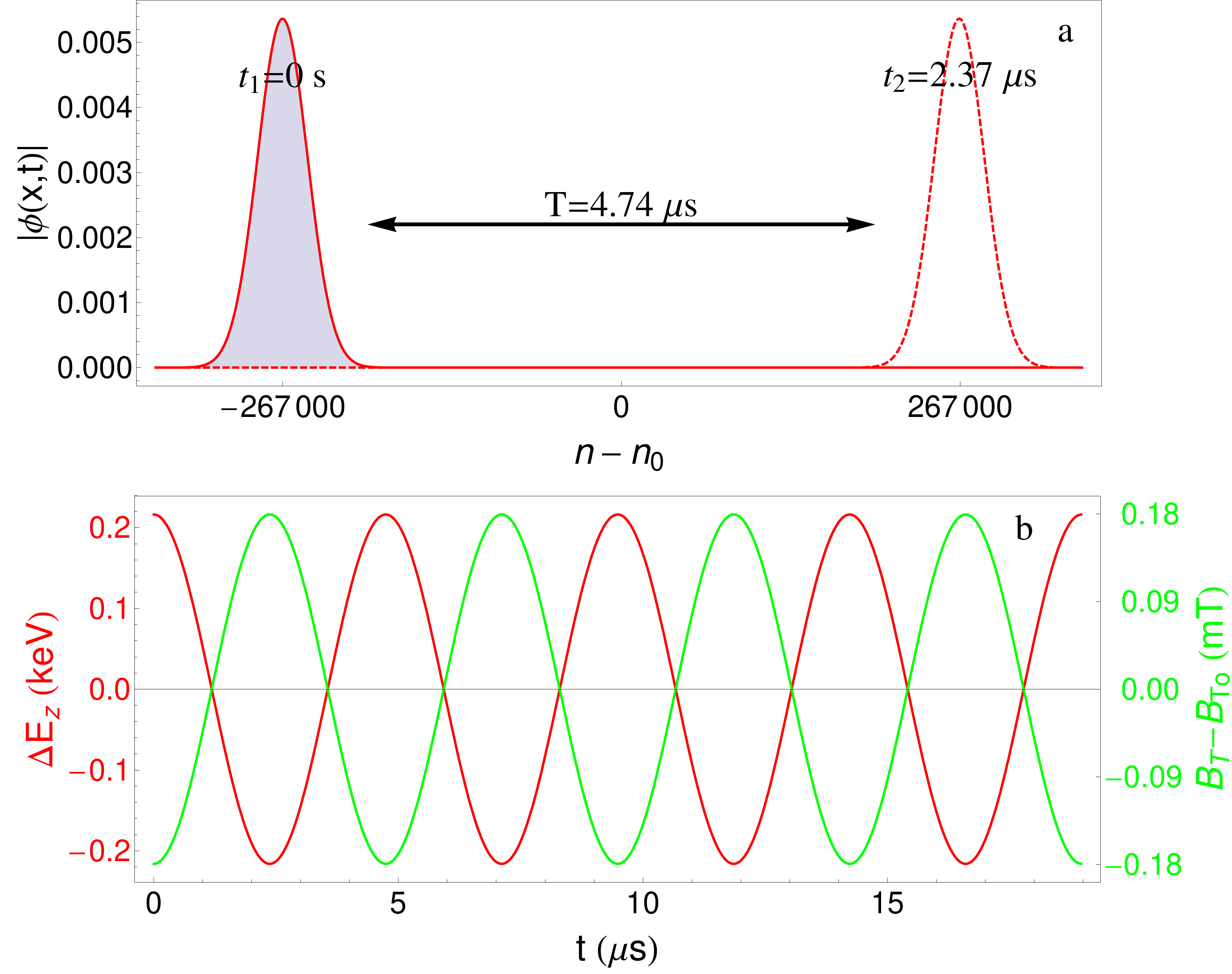}
\caption{(a) Amplitude of a coherent state of nanomagnet/photon system is shown as a function of photon number $n$. The large oscillations of this coherent state about $n_0=6.667\times 10^8$ occurs between $-267000$ (\textit{Filled}), and $+267000$ (\textit{Dashed}) in photon number with a period of $T=4.74 \mu s$. (b) Time evolution of the Zeeman energy of the nanomagnet (\textit{Red}), and transverse magnetic mode of the cavity field (\textit{Green}) at $z=-d$ are shown in this coherent state representation.}\label{fig:coherent}
\end{figure}

Coherent states, characterized by large oscillations in the photon number about $n_0$, can be expressed in the form
\begin{equation}
\phi(x,t)=\sum_{j=0}^{j_0}A_j e^{-i E_j t/\hbar} \psi_j(x),\label{12}
\end{equation}
where the phase factors $A_j$ are determined by setting $\phi(x,t=0)$ to a Gaussian wavefunction initially centered at $x_0=6.664\times 10^8$. Summation over the first 150 states is sufficient enough to obtain convergence in the dynamical properties. The coherent state oscillates with a period of 
$T=4.74$~$\mu$s, as shown in Fig.~\ref{fig:coherent}(a). The large oscillations in the collective dipole magnetic moment of the nanomagnet $\mu_z$, and in the cavity photon's magnetic field amplitude $\bm{B}_T$, are shown in Fig.~\ref{fig:coherent}(b) to emphasize the coherent nature of this state.

\begin{figure}[htp]
\centering
\includegraphics[width=8 cm]{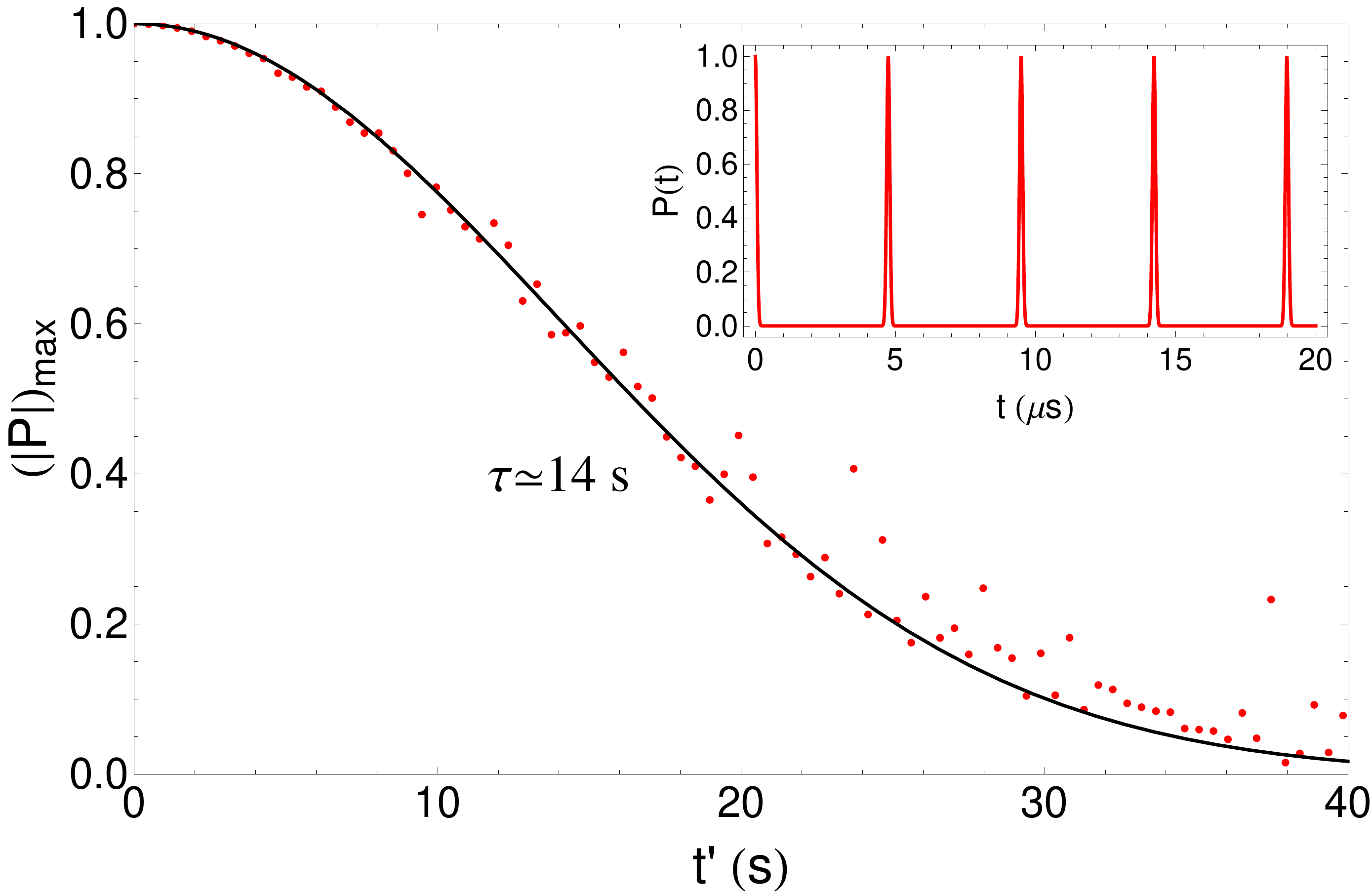}
\caption{Dephasing time of the coherent state is obtained by the Gaussian fit to the peak values of the dephasing function(Inset) at successive time intervals.}\label{fig:dephasing}
\end{figure}

The coherent properties of this nanomagnet-photon system will also depend on the dephasing of the coherent state $\phi(x,t)$, defined as the autocorrelation function, $P(t)=|\langle\phi(t)|\phi(0)\rangle|^2$. Exceptionally long dephasing times of roughly $\tau=14$ s are shown in Fig.~\ref{fig:dephasing}. Although this treatment is for zero temperature, the coherent properties of the nanomagnet-photon system should persist to as high a temperature (and over as long a timescale) as the macrospin description remains reliable. Recent work suggests that nanomagnet oscillators of approximately this size can be well-described by macrospins at room temperature\cite{Berkov2008}.
We have assumed an infinite $Q$ for the cavity, so the decoherence of the system is expected to be determined by photon leakage from the cavity, rather than these exceptionally long calculated times. 
We also find that other deviations from ideality for the nanomagnet, such as the cubic magnetic anisotropy terms, will only change the energy eigenvalues $E_j$ by $\ll 10^{-9}$~eV for the entire range of states involved in the oscillations described here, so such effects will not destroy or substantially reduce the coherent oscillations described here (although they may limit the dephasing times to shorter than shown in Fig.~\ref{fig:dephasing}).

The strong-field interactions between a nanomagnet of radius $100$ nm consisting of $10^9$ spins and a spherical microcavity roughly $1$ mm$^3$ in volume in the presence of a static magnetic field of $\sim 7$ T in magnitude indicate that strong-field coupling between magnets and spins is possible, and should substantially exceed the coupling observed in solids between orbital transitions and light. We find that the interaction Hamiltonian contains magnet-microwave mode coupling terms that can exceed several THz. Furthermore, the coherent states of our spin-photon coupling around the superradiance regime are characterized by large oscillations in photon number of the cavity (or equivalently the collective spin number of the nanomagnet) with exceptionally long dephasing times.

\begin{acknowledgements}
This work was supported by an ONR MURI.
\end{acknowledgements}

\end{document}